# The Sublimative Torques of Jupiter Family Comets and Mass Wasting Events on Their Nuclei

Jordan K. Steckloff* [a,b,c,d],
Nalin H. Samarasinha[a]

*corresponding author: jordan@psi.edu

[a]Planetary Science Institute, 1700 East Fort Lowell, Suite 106, Tucson, AZ 85719-2395, USA
[b]University Texas at Austin, Department of Aerospace Engineering and Engineering Mechanics, W.R. Woolrich Laboratories, C0600, 210 East 24th Street, Austin, TX 78712-1221 USA
[c]Massachusetts Institute of Technology, Department of Earth, Atmospheric and Planetary Sciences, Building 54-918, 77 Massachusetts Avenue, Cambridge, MA 02139-4307, USA
[d]Purdue University, Department of Earth, Atmospheric, and Planetary Sciences, 550 Lafayette St., West Lafayette, IN 47907, USA

## Abstract

Sublimative outgassing of comets produces torques that alter the rotation state of their nuclei. Recently, parameterized sublimative torque models have been developed to study rotation state changes of individual comet nuclei and populations of cometary bodies. However, these models simplify the interactions between the escaping gas and cometary surface into only a few parameters that hide the details of these complex interactions. Here we directly compare the X-parameter model (Samarasinha & Mueller, 2013) with the SYORP model (Steckloff & Jacobson, 2016) to tease out insights into the details of the gas-surface interactions driving sublimative torques. We find that, for both of these models to accurately model sublimative torques, the number of sublimating molecules that contribute to the net torque is largely independent of the detailed shape and activity of the nucleus, but rather depends primarily on the size of the nucleus and the effective heliocentric distance of the comet. We suggest that cometary activity must be largely restricted to regions of steep gravitational surface slopes (above the angle of repose), where mass wasting can refresh activity by shedding mantles of refractory materials and exposing fresh volatiles. We propose a new classification scheme for comets based on the frequency of this mass-wasting process (relative to the timescale of activity fading): quasi-equilibrium, episodic, quasi-dormant, and extinct.

1. **Introduction**

The sublimation of volatile species is a defining process of cometary bodies (Whipple 1950, 1951). These sublimating gases can blow off refractory dust, forming dust jets, gas jets, dust comae, gas comae, dust tails (Bessel, 1835), and synchronic dust bands (e.g., Kharchuk and Korsun, 2010). Volatile sublimation can also exert reaction pressures strong enough to fragment nuclei (Steckloff et al. 2015). Additionally, asymmetric sublimative gas emission can generate net torques on the nucleus that change the rotation state of the nucleus (e.g., Whipple, 1950; Samarasinha & Belton, 1995; Gutiérrez et al. 2003; Neistadt et al. 2003; Samarasinha & Mueller, 2013; Keller et al. 2015; Steckloff & Jacobson 2016), contribute to stria formation in the dust tail (Steckloff & Jacobson, 2016), induce avalanches and other mass wasting[1] events on the nucleus (Steckloff et al. 2016), or lead to disruption (e.g., Steckloff & Jacobson, 2016; Jewitt et al. 2016) or reconfigurations of the shape of the nucleus (Hirabayashi et al. 2016a). Traditional methods for studying such sublimative torques adopt an approach that integrate the torques caused by sublimation forces over the surface of the nucleus to compute the net torque (e.g., Gutiérrez et al. 2003; Neistadt et al 2003, Keller et al. 2015), and therefore require detailed information on the shape and activity of the nucleus. However, such sufficiently detailed information, especially the shape of the nucleus, is often only obtainable with high-resolution spacecraft observations of a comet nucleus, limiting the application of these methods to the handful of short-period comets that have been visited by spacecraft.

2. **Parameterized Sublimative Torque Models**

More recently, parameterized models of sublimative torques have been developed to study changes in the spin states of cometary bodies without high-resolution information. Samarasinha & Mueller (2013) developed a model of sublimative torques to describe the magnitude of a comet nucleus' orbitally averaged angular acceleration[2]:

$$\overline{\left|\frac{d\omega}{dt}\right|} = X \frac{2\pi \overline{Z(r_h)}}{R_n^2} \quad (1)$$

where $\omega$ is the angular velocity of the nucleus, $Z(r_h)$ is the H$_2$O flux at zero solar zenith angle as a function of heliocentric distance $(r_h)$, $R_n$ is the effective radius of the nucleus, and $X$ is a comet-specific constant that measures the specific change in rotational angular momentum averaged over the orbital period of a comet. $\overline{\left|\frac{d\omega}{dt}\right|}$ and $\overline{Z(r_h)}$ denote the orbitally averaged values during the active phase. Samarasinha & Mueller (2013), and later Mueller & Samarasinha (2018) used this parametric model to study the observed changes in the spin periods of comets 2P/Encke (Mueller et al. 2008), 9P/Tempel 1 (Belton et al. 2011; Chesley et al. 2013), 10P/Tempel 2 (Knight et al. 2011, 2012), 19P/Borrelly (Mueller & Samarasinha, 2015), 67P/Churyumov-Gerasimenko (Lowry et al. 2012; Mottola et al. 2014; Accomazzo et al. 2017), and 103P/Hartley 2 (Drahus et al. 2011; Belton et al. 2013; Knight et al. 2015). They discovered that this $X$ parameter is approximately constant amongst the sample of Jupiter family Comets (JFCs) considered in their study, varying by only a factor of a few in spite of active fractions of the comets' surfaces that varied by ~1.5 orders of magnitude (Samarasinha & Mueller, 2013; Mueller & Samarasinha, 2018).

---

[1] The term "mass wasting" refers to the downslope movement of geologic materials, which excludes volatile sublimation.

[2] Equation (1) is a rearrangement of the formalism presented in Samarasinha & Mueller (2013).

Steckloff & Jacobson (2016) describe another parameterized sublimative torque model analogous to the YORP effect (SYORP), which takes advantage of the similar manner in which photons and rarefied gas molecules are emitted from a porous regolith (e.g., Gombosi, 1994). The SYORP model breaks up the net sublimative momentum flux ($\phi_{sub}$) emitted from each area element ($dS$) of the surface of an icy body into two components with respect to a vector drawn from the body's center of mass to the area element: a radial component directed along this vector ($\phi_{rad}$) and a tangential component directed perpendicular to this vector ($\phi_{tan}$) (see Figure 1). Integrating over the surface ($S$) of the nucleus, the net sublimative momentum flux ($\Phi_{sub}$) and its radial ($\Phi_{rad}$) and tangential ($\Phi_{tan}$) components are

$$\Phi_{rad} = \iint_S \phi_{rad}\, dS \qquad (2)$$
$$\Phi_{tan} = \iint_S \phi_{tan}\, dS \qquad (3)$$
$$\Phi_{sub} = \iint_S \phi_{sub}\, dS = \iint_S \phi_{rad}\, dS + \iint_S \phi_{tan}\, dS \qquad (4)$$

Since the radial components ($\phi_{rad}$) [and thus, their integrated sum ($\Phi_{rad}$)] exert no torque on the nucleus, the SYORP model only considers the net tangential component of the sublimative momentum flux ($\phi_{tan}$) integrated over the surface of the nucleus ($\Phi_{tan}$).

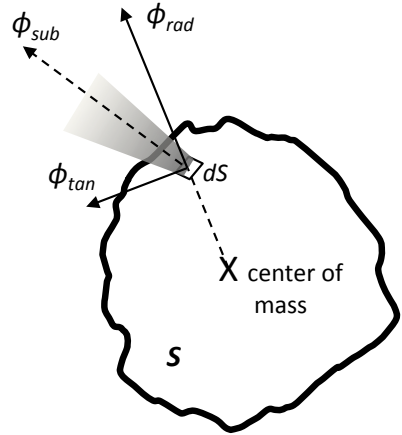

**Figure 1:** *SYORP divides sublimative momentum flux into components.* The reaction force of sublimating gas molecules on a cometary surface results due to momentum flux ($\phi_{sub}$) from each area element ($dS$) that is actively sublimating. This momentum flux can be split into radial ($\phi_{rad}$) and tangential ($\phi_{tan}$) components relative to the center of mass of the nucleus, of which only the tangential component exerts a torque on the body. The reaction forces are in the opposite sense to the momentum fluxes shown.

SYORP formalism modifies the YORP treatment presented in Rossi et al. (2009) for sublimative processes, resulting in a description of the average instantaneous magnitude of a comet nucleus's sublimative angular acceleration, normalized by instantaneous sublimative momentum flux at zero solar zenith angle

$$\overline{\left|\frac{d\omega}{dt}\right|} = \frac{3fP_S C_S}{4\pi\rho R_n^2} \quad (5)$$

where $f$ is the fraction of the surface are of the nucleus that is active (active fraction), $\rho$ is the bulk density of the body, $P_S$ is the sublimative momentum flux at zero solar zenith angle, and $C_S$ is the dimensionless SYORP coefficient, which accounts for orbitally averaged asymmetries in the emission of sublimating gas molecules from the nucleus (Steckloff & Jacobson, 2016). The SYORP coefficient is analogous to a body's YORP coefficient ($C_Y$) (see Rossi et al. 2009) in that it accounts for asymmetries in the shape of a body (Steckloff & Jacobson, 2016). Although one could integrate over the surface of a sublimating object to determine the net torque (the sum of the torques generated by each surface element of the nucleus; e.g., Nieshtadt et al. 2003; Keller et al. 2015; Hirabayashi et al. 2016), the SYORP model parameterizes these high-resolution details into a single coefficient that accounts for the average magnitude of the net torque acting on a comet nucleus over secular timescales (rather than instantaneous timescales).

The formulation of SYORP presented in Steckloff & Jacobson (2016) assumes that thermally primitive comets (e.g., dynamically new and dynamically young comets) have not built up a significant refractory (e.g., dusty) surface layer, resulting in ices exposed all over the surface ($f = 1$). This assumption does not hold for JFCs, which are thermally evolved comets with both active and inactive (non-sublimating) areas. Nevertheless, the SYORP effect still accounts for inactive regions of the nucleus through its SYORP coefficient[3]. Therefore, the SYORP coefficient ($C_S$) can be more accurately thought of as the fraction of a nucleus' total sublimative momentum flux that generates a net sublimative torque. Although the SYORP model was developed assuming negligible outgassing from the unilluminated side of the comet, the SYORP coefficient can account for some night-side outgassing, so long as the illuminated hemisphere dominates the comet's overall gas flux. If we define $\langle \phi_{sub} \rangle$ as the average net sublimative momentum flux and $\langle \phi_{tan} \rangle$ as its average net tangential component

$$\langle \phi_{tan} \rangle = \frac{\Phi_{tan}}{S} \quad (6)$$

$$\langle \phi_{sub} \rangle = \frac{\Phi_{sub}}{S} \quad (7)$$

then the SYORP coefficient ($C_S$) and active fraction ($f$) can be defined as

$$C_S = \frac{\Phi_{tan}}{\Phi_{sub}} = \frac{\langle \phi_{tan} \rangle}{\langle \phi_{sub} \rangle} \quad (8)$$

$$f \approx \frac{\Phi_{sub}}{\iint_S P_S \mathbf{n} \cdot d\mathbf{S}} = \frac{k\langle \phi_{sub} \rangle}{P_S} \quad (9)$$

where $\mathbf{n}$ is a unit vector pointing toward the Sun, $d\mathbf{S}$ is a unit vector normal to the surface area element, and $k$ is the ratio of the nucleus's illuminated surface area to its cross-sectional area relative to zero solar zenith angle (i.e., surface area of the plane defined by the terminator), averaged over the portion of the comet's orbit during which its surface is actively sublimating.

---

[3] This has no analogy in the YORP effect because the entire surface can reradiate thermal photons.

Thus, $f$ is approximately equal to the ratio of the measured total momentum flux to the maximum possible momentum flux (were the entire surface covered in ice). Geometric considerations show that $k \approx 2$ for an approximately spheroidal nucleus or a bilobate nucleus in which both lobes are approximately spherical with non-zero spin pole obliquity[4]. Although this is consistent with models of the shapes of JFC nuclei observed by spacecraft (e.g., Farnham et al. 2005; Farnham & Thomas, 2013a, 2013b; Jorda et al. 2015), we will later show that the exact shape of the nucleus is largely unimportant. Equation (5) can be combined with equations (8) and (9)

$$\overline{\left|\frac{d\omega}{dt}\right|} = \frac{3k\langle\phi_{tan}\rangle}{4\pi\rho R_n^2} \qquad (10)$$

Additionally, as the average tangential sublimative flux ($\langle\phi_{tan}\rangle$) is a function of heliocentric distance

$$\langle\phi_{tan}\rangle = f_{tan}\frac{\overline{Z(r_h)}}{k}m_{water}\langle v_{water}(r_h)\rangle \qquad (11)$$

where $f_{tan}$ is the effective tangential fraction of the theoretical volatile flux at zero solar phase angle that contributes to a net torque, $\frac{\overline{Z(r_h)}}{k}$ is the average water production rate (molecules per unit area per unit time), $m_{water}$ is the mass of a water molecule, and $\langle v_{water}(r_h)\rangle$ is the average molecular outflow velocity of the sublimating water molecules in the direction normal to the surface. If we assume the sublimating gas to be in thermal equilibrium with the ice from whence it originated, then the molecular outflow velocity is equal to the molecular thermal velocity, which is dependent upon heliocentric distance.

By combining equations (10), and (11), we derive

$$\overline{\left|\frac{d\omega}{dt}\right|} = \frac{3f_{tan}\overline{Z(r_h)}m_{water}\langle v_{water}(r_h)\rangle}{4\pi\rho R_n^2} \qquad (12)$$

While other sublimative torque models have been developed (e.g., Jewitt, 1997; Keller et al. 2015; Hirabayashi et al. 2016a), they will not be discussed in this work, as they are not parameterized in the same manner as the X-parameter and SYORP models. This work instead will focus on comparing the Samarasinha & Mueller (2013) and Steckloff & Jacobson (2016) parametric sublimative torque models, which must be consistent with one another for both to be correct.

We may therefore compare equation (12) with equation (1) and obtain

$$\frac{3f_{tan}\overline{Z(r_h)}m_{water}\langle v_{water}(r_h)\rangle}{4\pi\rho R_n^2} = X\frac{2\pi\,\overline{Z(r_h)}}{R_n^2} \qquad (13)$$

---

[4] Low spin pole obliquity would result in each lobe shadowing the other during a significant faction of the comet's orbit, potentially increasing $k$ by up to a factor of two.

and solve for $f_{tan}$

$$f_{tan} = X \frac{8\pi^2 \rho}{3 m_{water} \langle v_{water}(r_h) \rangle}  \quad (14)$$

Both the parametric models of Samarasinha & Mueller (2013) and Steckloff & Jacobson (2016) agree that sublimative torques on comet nuclei are only significant in regions of the Solar System where sublimative cooling is the dominant heat-loss mechanism (inward of ~2 AU, ~8 AU, and ~100 AU for $H_2O$, $CO_2$, and CO respectively). In the region of space where these volatiles vigorously sublimate (where sublimative cooling is the dominant heat-loss mechanism), their average molecular thermal velocity ($\langle v(r_h) \rangle$) varies by less than ~10% (see Figure 2). We can therefore consider $\langle v_{water}(r_h) \rangle$ in equation (14) to be approximately constant to first order.

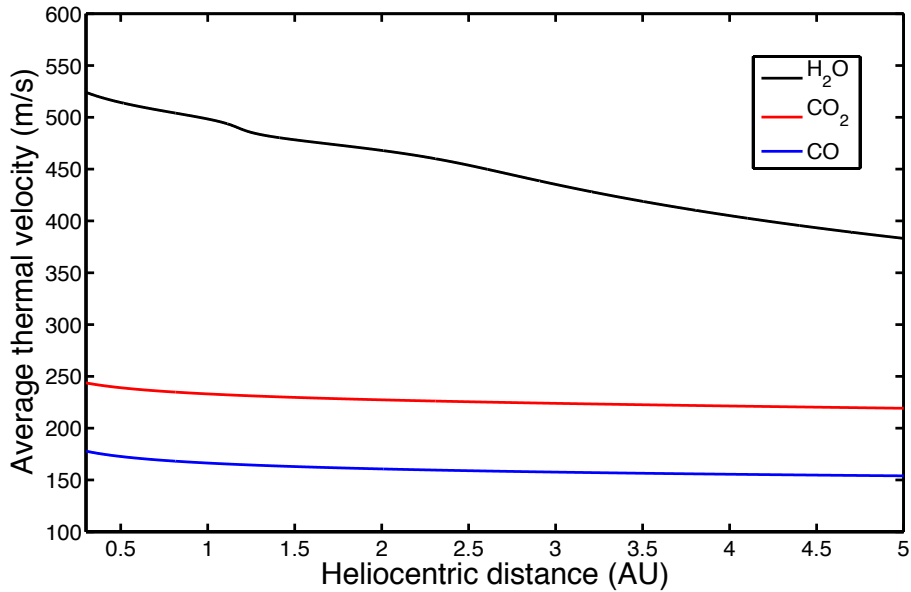

**Figure 2:** *Heliocentric distance dependence of average sublimation thermal velocity.* The average magnitude of the velocity of the escaping gas-phase molecules depends on the temperature of the sublimating volatiles. Most Jupiter Family Comets (JFCs) orbit the Sun between 0.3 and 10 AU. However, sublimative torques are only significant when volatiles are vigorously sublimating (inward of ~2 AU, ~8 AU, and ~100 AU for $H_2O$, $CO_2$, and CO respectively). The plot above shows that the average magnitude of the thermal velocity of the three most common cometary volatiles ($H_2O$, $CO_2$, and CO) in the region of the JFCs and where they vigorously sublimate varies by less than ~10%. We can therefore treat the thermal velocity of the sublimating gas as though it were effectively constant. Thermal velocity was computed using the numerical method of Steckloff et al. (2015).

Additionally, the estimated densities of JFCs range cover a narrow range. Spacecraft observations allowed for the direct measurement of the gravity field of comets 9P/Tempel 1 and 67P/Churyumov-Gerasimenko. Observations of the ejecta plume evolution during the Deep Impact experiment suggest that 9P/Tempel 1's density is 400 ± 200 kg m$^{-3}$ (Richardson et al.

2007), while orbital changes of the Rosetta spacecraft suggest that 67P/Churyumov-Gerasimenko's density is 533 ± 6 kg m$^{-3}$ (Pätzold et al. 2016). These two measurements are consistent with one another, and suggest that the densities of JFC nuclei occupy a narrow range of values. Furthermore, the density of comet 103P/Hartley 2 was estimated to be 300 ± 100 kg m$^{-3}$ by assuming that the "neck" of the nucleus lies along an equipotential surface (Thomas et al. 2013). While this estimate is less reliable than direct measurements, it nevertheless is consistent with JFC densities occupying a narrow range. Furthermore, the densities of JFCs 19P/Borrelly and 81P/Wild 2 have been estimated from their non-gravitational dynamical behavior to be 180-300 (Davidsson & Gutiérrez, 2004) or 290 – 830 kg m$^{-3}$ (Farnham & Cochran, 2002) for 19P/Borrelly, and 600-800 kg m$^{-3}$ (Davidsson & Gutiérrez, 2006) for 81P/Wild 2. Additionally, Asphaug and Benz (1996) used dynamical considerations to estimate the density of Comet Shoemaker-Levy 9 to be ~600 kg m$^{-3}$ (Asphaug & Benz, 1996). Although these three estimates are based on model assumptions that likely have significantly greater errors than estimates derived from direct observations of the nucleus, they nevertheless are roughly consistent with the densities of JFCs occupying a narrow range of values. We therefore treat the density of JFC nuclei ($\rho$) of equation (14) as approximately constant.

Finally, Samarasinha & Mueller (2013) and Mueller & Samarasinha (2018) found that the value of *X* varied by less than a factor of a few across the comets they considered. Therefore, the entire right-hand side of equation (13) is constant to within a small factor (within half an order of magnitude), leading us to conclude that

$$f_{tan} \sim constant \qquad (14)$$

This means that the number of molecules that contribute to the net torque of the nucleus depends *primarily* on the size and heliocentric distance of the nucleus, and is *largely independent* of the detailed shape, activity distribution, and active fraction of the nucleus.

Recent observations indicate that comet 41P/Tuttle-Giacobini-Kresák (41P/TGK) experienced sublimative torques that increased it spin period from 20 hours to more than 46 hours in approximately two months (Farnham et al. 2017; Knight et al. 2017; Schleicher et al. 2017; Bodewits et al. 2018). However, we exclude 41P/TGK from our study since the size of its nucleus is not yet well determined, and the *X* parameter and SYORP coefficient associated with 41P/TGK's spin down are not presently known.

3. **Discussion**

Samarasinha & Mueller (2013) considered the effect of stochastic cancellations of torque for comets with different levels of activity. They argue that, because more active comets are more likely to possess active areas that produce opposing torques, a more active nucleus should not necessarily experience a net torque of greater magnitude than a nucleus with less activity. Since sublimative torques result from the net tangential momentum flux, both the X parameter and $f_{tan}$ should be largely independent of a comet nucleus' active fraction *f*. Such torque cancellations require that active areas of the nucleus to produce torques that are directed somewhat randomly and have strengths of comparable magnitudes.

Nevertheless, torque cancellations may not be the only reason why young comets with activity and evolved comets with activity have nearly the same X parameter and $f_{tan}$. Jupiter Family Comets (JFCs) have a dynamical lifetime of ~300,000 years (Duncan et al. 2004), during which the surface of the nucleus is expected to thermally evolve. $CO_2$ sublimates vigorously inward of ~8 AU from the Sun, and $H_2O$ sublimates vigorously inward of ~2 AU (Steckloff et al. 2015a; Steckloff & Jacobson, 2016). Thus, a comet nucleus's volatile sublimation fronts from $CO_2$ and $H_2O$ will recede below the surface of the nucleus inwards of ~8 and ~2 AU from the Sun respectively (assuming that the $CO_2$ and $H_2O$ molecules are not intimately mixed at the molecular level or forming a clathrate). This leaves a volatile poor, refractory surface above the ice that acts as a very effective thermal insulator due to its low thermal inertia (Gulkis et al., 2015; Davidsson et al., 2013; Groussin et al., 2013; Lisse et al., 2005; Lamy et al., 2008) that significantly reduces ice sublimation and sublimative torques. Thus, comets with younger active areas such as 103P/Hartley 2 (Steckloff et al. 2016) are expected to generate greater sublimative torques per unit surface area and would therefore be expected to have a higher X parameter and $f_{tan}$ than comets with older active areas such as 9P/Tempel 1 (Yeomans et al. 2005). As this is not the case, this suggests that a process may exist for reactivating comet nuclei such that $f_{tan}$ is independent of the thermal age of the comet.

We consider whether mass wasting processes could contribute to the independence of the X parameter and $f_{tan}$. We classify the gravitational slope[5] of the comet nucleus' surface elements as either steep-sloped or shallow-sloped, with the angle of repose of the surface material marking the boundary between these two scenarios. The angle of repose of a granular material denotes its maximum stable surface slope angle, above which its surface layers are unstable and prone to mass wasting. The angle of repose for most non-cohesive granular materials is 30 to 45° (Lambe & Whitman, 1969), consistent with the observed loose materials on cometary surfaces (Groussin et al. 2015; Steckloff et al. 2016; Vincent et al. 2016a). However, steeper slope angles can be stable if the material has cohesive strength (which increases the material's angle of repose), consistent with observations of comet nuclei (Bowling et al. 2014; Groussin et al. 2015; Steckloff et al. 2015). Because steep surfaces of comet 67P/Churyumov-Gerasimenko and 81P/Wild 2 are active and therefore volatile rich (Sekanina et al. 2004; Vincent et al. 2015; Vincent et al. 2016a), it is reasonable to assume that the presence of volatile ices increases the cohesive strength of cometary material. Alternatively, fracturing of steep surfaces on comets, perhaps due to thermal fatigue (El-Maarry et al. 2015a), may reduce the cohesion of cometary materials and facilitate slope failure (Vincent et al. 2016a). In either case, steeper slopes (above the angle of repose) should become less cohesive the longer they are exposed at the surface, and prone to mass wasting, consistent with observations (Vincent et al. 2016a, 2016b).

---

[5] The gravitational field at the surface of such small, irregular object can be highly variable in both direction and magnitude. All slope angles here are relative to the local gravitational equipotential surface, which forms a curved reference surface normal to the local gravitational acceleration vector (a "flat" surface).

Gas molecules leaving a nucleus are ejected on average perpendicular to the surface of their source. *Because steeper slopes are more likely to have an orientation that deviates significantly from the direction of a vector drawn from the center of mass of the nucleus and the surface element, steeper slopes are likely largely responsible for generating the net tangential momentum flux $\Phi_{tan}$.* We imagine that both steep-sloped and shallow-sloped surfaces are initially a homogeneous primordial mixture of refractory materials and volatile ices. As solar insolation warms their surfaces, the ice begins to sublimate, driving activity. Over time, the ice should recede below the surface, reducing cohesion of the surface layer and forming an insulating lag deposit (i.e., mantle of refractory materials), which may also further thicken through deposition of material ejected from other parts of the nucleus (Thomas et al. 2015), diminishing cometary activity over time. Because this mantle of refractory materials covering shallow slopes is stable from avalanches, activity on these surface elements should gradually shut down over time as the mantle grows (e.g., Brin & Mendis, 1979; Fanale & Salvail, 1984; Prialnik & Bar-Nun, 1988; Rosenberg & Prialnik, 2009). Some sublimative activity may be seen originating from these shallow-sloped surfaces due to formation of surface frosts during cometary night. This surface frost formation could be either due to re-condensation of inner coma water molecules on the colder night-side surface (Sunshine et al. 2006; De Sanctis et al. 2015) or due to the re-condensation of sublimating water molecules from sub-surface ices as they reach the colder surface of the nucleus during the night (De Sanctis et al. 2015). Such surface frosts would ultimately be removed from the comet through various physical processes (e.g., sublimation during the day time, solar wind sputtering, gas collisions, photolysis), and therefore would need to be replenished from a primary source of volatiles. In contrast, the refractory mantles covering steep slopes are unstable and will readily avalanche, which can excavate down to ice rich layers and reactivate the steeper slopes (see Figure 3).

Such a volatile-driven scarp retreat process was first proposed by Britt et al. (2004) to explain the formation of mesas on comet 19P/Borrelly. This study suggested that mass-wasting events would remove non-volatile materials from steep slopes, allowing scarps to retreat through sublimation. More recently, the *Rosetta* spacecraft observed consolidated, cliff-like structures on 67P/Churyumov-Gerasimenko that appear to desiccate through outgassing, forming fractured cliff-like surfaces that eventually collapse (e.g., El-Maarry et al. 2015b). The collapsed cliff would re-expose volatile-rich materials, allowing activity to continue from these steep terrains (Vincent et al. 2016a, 2016b). The resulting debris may form a talus field, or it may be ejected via sublimation, and never reach the surface (Vincent et al. 2016a, 2016b; Steckloff & Melosh, 2016). In any event, mass-wasting processes provide a natural mechanism for refreshing cometary activity on steeper slopes while shutting down activity on shallower surfaces as a comet thermally ages. The resulting persistence of steeper sources of activity, thus naturally explains the independence of the X parameter and $f_{tan}$ with the thermal age of the comet.

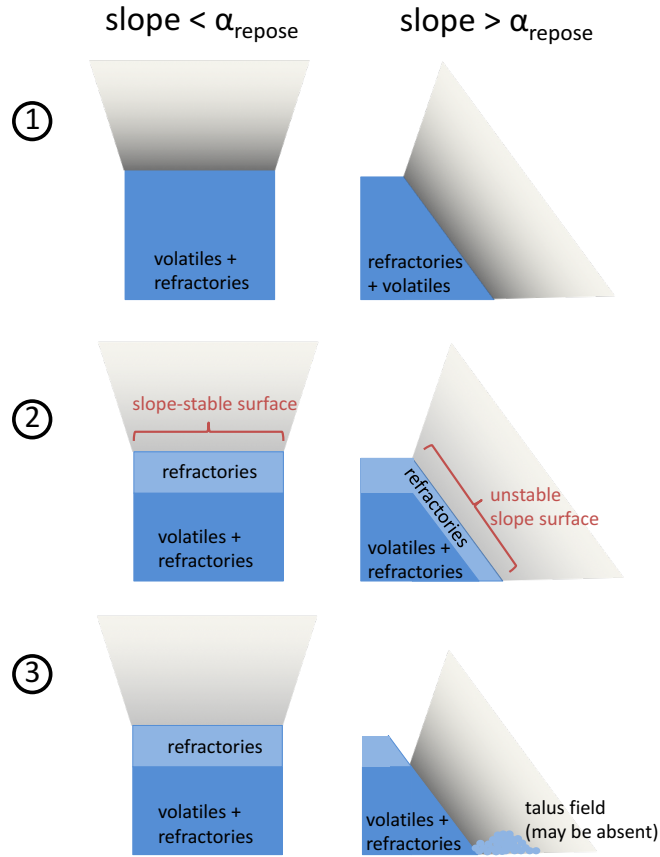

**Figure 3:** *Evolution of sublimating surfaces of comet nuclei over time.* The active surfaces of a comet nucleus follow one of two slope angle-dependent evolutionary paths. The difference between these two paths is the slope of the surface relative to its angle of repose ($\alpha_{repose}$). Although different terrains may have different angles of repose, shallower slopes are less than the angle of repose (surface slope $< \alpha_{repose}$) and steeper surfaces having slope angles greater than the angle of repose (surface slope $> \alpha_{repose}$). (1) Both a shallowly and steeply sloped (relative to the angle of repose) regions with relatively pristine material should exhibit sublimative cometary activity (shown in gray) when illuminated by the Sun. (2) Refractory materials will accumulate on the sublimating surfaces over time, forming a thermally insulating refractory mantel that significantly diminishes cometary activity. While this refractory mantel is stable on the shallower surfaces, it is unstable on steeper surfaces and prone to mass-wasting. (3) Mass-wasting on the steeper slopes excavates down to buried ices, reactivating the steeper surfaces. Darker gray indicates strong outgassing while diminished outgassing is represented by light gray.

Additionally, the persistence of steeper sources of activity implies that observations of Jupiter Family Comets should preferentially find jets and outgassing associated with steeply sloping surfaces (e.g. scarp faces and cliffs) and show evidence of mass wasting or avalanches. While a mass wasting event could produce a short-live outburst during the short duration of the event (Steckloff & Melosh, 2016), the ice-rich talus fields (debris deposit) below the steeper terrains (Vincent et al. 2016a), and possibly newly exposed ices on the slopes themselves (Steckloff et al. 2016) would produce long-lasting cometary activity. Farnham et al. (2013)

detected 11 jets in the Stardust-NExT images of comet 9P/Tempel 1, seven of which could be traced to observed and illuminated portions of the nucleus. The likely source region of all seven jets are bisected by a long, terraced scarp, which Farnham et al. (2013) interpreted as the likely source of the jets. Sekanina et al. (2004) found that the jets on comet 81P/Wild 2 originate predominantly from steeply sloped source regions. On the theoretical front, Steckloff et al. (2016) argue that comet 103P/Hartley 2 rotated at a faster rate in the recent past, such that the tip of its small lobe was a steep scarp that experienced an avalanche, activating the dominant activity on the tip of Hartley 2's small lobe. Furthermore, much activity in the northern hemisphere of comet 67P/Churyumov-Gerasimenko originates from fractured cliff faces with talus deposits at their bases (Vincent et al. 2016a) and from the walls of active pits (Vincent et al. 2015). 67P/Churyumov-Gerasimenko's outbursts have been traced to the faces of scarps (Steckloff & Melosh, 2016; Grün et al. 2016; Pajola et al. 2017). Furthermore, talus is present at the bases of many cliffs on comets 81P/Wild 2 (Brownlee et al. 2004) and 67P/Churyumov-Gerasimenko (e.g., El-Maarry et al. 2015b; Vincent et al. 2016a; Pajola et al. 2017, El-Maarry et al. 2017), consistent with the nature of the mass wasting processes. Since mass wasting-driven outbursts are inherently episodic, they would be restricted to Type III outbursts in the Belton (2010) taxonomy of outbursts.

Recent spacecraft missions have found that not all sublimative activity on comets is clearly associated with a steeply sloping surface. Water ice frosts were observed on the surfaces of both comets 9P/Tempel 1 (Sunshine et al. 2006) and 67P/Churyumov-Gerasimenko (De Sanctis et al. 2015). However, such frosts are thought to be the product of water molecules condensing out of the inner coma and onto the surface of the nucleus rather than primary outcrops of water ice (Sunshine et al. 2006; De Sanctis et al. 2015). Alternatively, such frosts could form from sublimating subsurface water ice that recondenses onto the colder night-time surface of the nucleus (De Sanctis et al. 2015), which is an expected phenomenon as ice recedes below the surface. Relatedly, the smooth waist terrain of comet 103P/Hartley 2 is a highly active (A'Hearn et al. 2011; Feaga et al. 2017), shallowly sloping terrain far from a steeply sloping surface (Steckloff et al. 2016). Similarly, the Imhotep terrain of comet 67P/Churyumov-Gerasimenko is a smooth, flat terrain that is also active (e.g., Marshall et al. 2017, El-Maarry et al. 2017). However, both of these terrains are likely formed from the ballistic deposition of ice-rich materials ejected from other parts of the nucleus (A'Hearn et al. 2011; Keller et al. 2017; Lai et al. 2017), and are therefore symptomatic of activity elsewhere on the nucleus. 103P/Hartley 2's smooth waist in particular appears to be formed from ice-rich chunks ejected from Hartley 2's dominant small lobe active area (Hirabayashi et al. 2016b), which was recently activated by a mass wasting event (Steckloff et al. 2016). Therefore, even activity from seemingly smooth, flat terrains may also be the result of mass wasting events on steep terrains.

*Mass wasting events may produce fine layering*

Such mass wasting-driven activity could produce significant layering over time, as debris fields overlap and overlay one another. Significant layering has been observed on multiple comet nuclei: 9P/Tempel 1 (A'Hearn et al. 2005; Thomas et al. 2006; Belton et al. 2007), 67P/Churyumov-Gerasimenko (Massironi et al. 2015; El-Maarry et al. 2015b), and possibly 81P/Wild 2 (Brownlee et al. 2004). This is not to suggest that mass wasting is the only mechanism for layering in comet nuclei, as formational processes (Belton et al. 2007; Massironi et al. 2015), volatile outgassing (Mousis et al. 2015), or nucleus reconfiguration cycles (Hirabayashi et al. 2016a) may also cause significant layering of comet nuclei. The layering by mass wasting is expected to be limited to specific regions on the nucleus or steep (or formerly steep) terrains, rather than being global in nature.

*Dormancy, Reactivation, and long-term survival of JFCs*

The long-held notion that comet nuclei slowly sublimate away or become extinct is an oversimplification that doesn't account for potential reactivation. Comet nuclei should have the potential to remain active so long as there are steep slopes available to undergo mass wasting. While such a comet may temporarily exhibit little activity, an onset of a mass wasting event could reactivate the nucleus by exposing volatile ice to the surface. Conversely, a comet's activity should rapidly fade away if its nucleus were to reach a shape that lacks steep slopes, as such a shape could no longer reactivate through mass wasting events. Such a comet would ultimately appear dormant. However, surface slope angles of comet nuclei can be very sensitive to changes in rotation state (i.e., Steckloff et al. 2015), allowing sublimative torques to change the slope angles, and potentially the slope stability, of a comet's surface. While low levels of sublimation could still be present on a dormant nucleus, the level of activity would be significantly less than that of a typical, active comet. If such low-level sublimative activity were insufficient to change the rotation state of the nucleus (and thus the surface slope angles of the nucleus), then the comet will remain dormant and effectively become an extinct comet. The largely smooth surface of 9P/Tempel 1 suggests that this comet may be evolving toward dormancy, and possibly extinction.

However, it is also possible that such low-level sublimative activity could change the nucleus rotation state significantly enough to change some shallow surface slopes of the nucleus into steep slopes, especially if the nucleus is small, as small nuclei will require comparatively small torques to change the rotation state. In such a case, the surface would experience an avalanche that may excavate down to ice-rich material, reactivating the nucleus' cometary activity. Steckloff et al. (2016) proposed that such an event occurred on comet 103P/Hartley 2 between 1984 and 1991, and showed that Hartley 2's observed activity is strongly suggestive of such an event. This suggests that comet 103P/Hartley 2 is probably the first known comet to be reactivated through mass wasting.

Other mechanisms have been previously proposed for activating/reactivating comet nuclei. Dynamical perturbations can lower the perihelia of JFCs, bringing them closer to the sun. This would increase the amplitude of the orbital thermal wave that conducts into the comet each orbit, which could allow enough solar heat to reach deeper ices to reactivate the comet (e.g., Ferrín et al. 2013). However, such a mechanism would not explain the sustained activity of comets "decoupled" from gravitational perturbations with a giant planet, such as comets 2P/Encke (Levison et al. 2006) and 107P/Wilson-Harrington (Fernández et al. 2002). Collisions with asteroids and meteoroids have also been proposed as a mechanism to excavate through cometary mantles and expose ices (e.g., Jewitt et al. 2010; Haghighipour et al. 2016). However, the dearth of impact craters on the surfaces of JFC nuclei visited by spacecraft suggests that collisional activation is unlikely to contribute significantly to their activation. Finally, sublimative torques may cause stresses to build within comets that cause their nuclei to split or disrupt, exposing buried parts of the nucleus to the sun and causing cometary activity (e.g., Jewitt et al. 2010, 2014, 2017; Steckloff & Jacobson, 2016, Hirabayashi et al. 2016a; El-Maarry et al. 2017). However, rotational bursting is a violent process that may also lead to a comet's demise, and produces daughter nuclei that are clearly dynamically related. By contrast, mass wasting is a gentle process that can affect all comet nuclei. Therefore, mass wasting-driven reactivation may be cyclical, and a quintessential component of comet evolution.

However, different nucleus shapes and the "evolutionary ages" of a comet's sublimative activity may determine the interval between reactivation (i.e., mass wasting) events, leading to three distinct regimes of surface reactivation: *quasi-equilibrium, episodic, and quasi-dormant*. Young comets with many regions of steep slopes may experience regular episodes of mass-wasting on timescales shorter than the fading timescale of an individual active area. This would lead to a relatively constant level of activity between apparitions (quasi-equilibrium activation), consistent with observations of comets 81P/Wild 2 and 67P/Churyumov-Gerasimenko (e.g., Bertini et al, 2010; Vincent et al. 2013). Because mass-wasting events appear to be correlated with outburst activity (Steckloff & Melosh, 2016), such comets would likely experience significant outbursts, also consistent with observations (e.g., Vincent et al. 2016b). Comet 1P/Halley may also be such a quasi-equilibrium comet, owing to its consistent activity over two millennia (e.g., Yeomans & Kiang, 1981), with rugged terrain on its nucleus (Keller et al. 1987).

As a comet ages and begins to run out of scarps susceptible to mass-wasting, the interval between reactivation events may become comparable to, or longer than, the fading timescale of an individual active area, each mass-wasting event would appear to correlate with a dramatic brightening of the comet. This would lead to long lived episodes of cometary activity that fade over secular timescales (episodic activation). The avalanche-outburst model of Steckloff & Melosh (2016) suggests that any such episodic comets would likely exhibit little to no outburst activity. The fading timescale of cometary activity is currently unknown, but likely longer than many orbits. Thus, 103P/Hartley 2 may be an ideal case to study the secular fading of cometary activity over time if the avalanche model of Steckloff et al. (2016) on Hartley 2's reactivation proves correct. However, if the activity fading timescale is on the order of dozens of orbits, Hartley 2's activity may take a very long time to appreciably fade, as recent studies (e.g., Knight & Schleicher, 2013) may indicate. In such a case, the apparent constant activity of comet 1P/Halley (Yeomans & Kiang, 1981) may simply be indicating that a large number of orbits are required to significantly affect sublimative outgassing rates. The apparent hyperactivity of comet 252P/LINEAR's 2016 apparition relative to its 2000 apparition (Li et al. 2017) suggests that 252P/LINEAR may be in the episodic regime.

In addition, old comets may experience such long intervals between reactivation events (significantly longer than their fading timescale), that they only experience episodes of activity a small fraction of the time. Such long intervals could significantly extend the sublimative lifetime of a comet, allowing its orbit to dynamically evolve over orders of magnitude more time than if the comet were sublimating constantly. For example, comet 2P/Encke likely required ~80,000 perihelion passages to evolve into its current orbit, which is ~200 time longer than the expected sublimative lifetime for a Jupiter Family Comet of ~400 perihelion passages (Levison et al. 2006). This suggests that Comet 2P/Encke has been dormant for the vast majority of its dynamical life (Levison et al. 2006). Our model suggests that a mass wasting event(s) may have triggered the episode of activity that we see today. An observer of a random member of an Encke-like comet population would most likely see a dormant object (quasi-dormant regime of activation) that appears asteroidal, and would only identify the object's cometary nature if it were experiencing an episode of activity. Therefore, one would expect many dormant Encke-like objects for every active Encke-type comet, consistent with previous studies (Asher et al. 1993; Levison et al. 2006). Therefore, 2P/Encke may be a quasi-dormant regime comet that is currently in an active phase of its life, and that other Encke-like objects may be dormant, rather than extinct comets (Levison et al. 2006). Indeed, Ye et al. (2016), which surveyed meteor-producing dust trails and the Near-Earth Object (NEO) population, found that at least $2.0 \pm 1.7$ percent of the NEO population are dormant comets.

Finally, comets are not likely to remain permanently within a single reactivation regime. Rather, the interval between mass wasting events is likely to increase over time as steep slopes erode and flatten, consistent with the surface evolution model of Vincent et al. (2017). We propose a new general paradigm of cometary activity evolution in which comets transition from one reactivation regime to another. Young comets likely begin in the quasi-equilibrium regime, before transitioning to the episodic regime as the interval between mass wasting events lengthens. Eventually, these comets would evolve into the quasi-dormant regime, before ultimately becoming an extinct comet. *Because this evolutionary scheme looks at the mechanism required for refreshing cometary activity (rather than the activity itself), it may be a more useful scheme for classifying the evolutionary ages of comet nuclei.* According to this scheme, comets 9P/Tempel 1, 81P/Wild 2, and 67P/Churyumov-Gerasimenko, all of which exhibit regular outbursts (A'Hearn et al. 2005; Bertini et al. 2012; Tubiana et al. 2015; Pajola et al. 2017), are quasi-equilibrium comets and evolutionarily young. Comet 103P/Hartley 2 would be an episodic comet due to its lack of reported outbursts and suspected recent reactivation event (Steckloff et al. 2016). This somewhat surprising classification shows 103P/Hartley 2 to be an evolutionarily older comet than other JFCs visited by spacecraft, in spite of its very high rate of activity. Thus, while this scheme does not rule out a general trend of decreasing active fraction with increasing evolutionary age over secular timescales, the active fraction is likely a poor indicator of a comet's evolutionary age.

Our analysis does not place strong constraints on the relative evolutionary ages of these three comets (9P/Tempel 1, 81P/Wild 2, and 67P/Churyumov-Gerasimenko) within the quasi-equilibrium regime. However, Vincent et al. (2017) used topographic evolution arguments to suggest that 9P/Tempel 1 is evolutionarily older than either 81P/Wild 2, and 67P/Churyumov-Gerasimenko. Since steep topographic variation is needed for mass-wasting to occur, our analysis is consistent with this interpretation.

*Main-Belt Comets*

This evolutionary scheme assumes that sublimative torques dominate the rotational evolution and reactivation of cometary nuclei. This assumption holds well for ecliptic comets (e.g., JFCs and Encke-type) and some nearly isotropic comets (e.g., Halley-type comets). However, this scheme may not be applicable to the main-belt comets (MBCs), for which other physical processes become significant. Whereas the short dynamical lifetimes of ecliptic comets restrict their spin evolution to the effects of sublimative torques, MBCs reside in the relatively dynamically stable Main Asteroid belt (e.g., Dones et al. 2015). Such dynamical stability allows an otherwise dormant or extinct nucleus to slowly spin up not just through sublimative torques, but also through the YORP effect, potentially reactivating a nucleus through radiative forces. Indeed, MBCs (62412) 2000 SY178 and 133P/Elst-Pizarro are rotating close to their spin disruption limit, with spin periods of 3.33±0.01 hr (Sheppard & Trujillo, 2015) and 3.471±0.001 hr (Hsieh et al. 2004) respectively. These extreme rotation rates can potentially trigger mass wasting events that unearth buried volatile ices. This is consistent with some MBC activity being driven by the sublimation of volatile ices (Hsieh et al. 2015), rather than through rotational shedding of material. Additionally, Jewitt et al. (2010) and Haghighipour et al. (2016) have proposed that impact events may trigger MBC activity by excavating craters that uncover buried volatile ices. Thus, the collisional nature of the Main Asteroid Belt may provide yet another mechanism for triggering MBC activity, further diverging from the assumptions of our evolutionary model.

4. **Conclusions**

We compare two parametric sublimative torque frameworks, the X-parameter model of Samarasinha & Mueller (2013) with the SYORP framework of Steckloff & Jacobson (2016). For both of these frameworks to accurately model the sublimative torques experienced by cometary bodies, the fraction of outflowing molecules that contribute to a net torque of the nucleus ($f_{tan}$) must be an approximately constant fraction of the average production rate for all comets. This entails that the net sublimative torque experienced by a comet nucleus depends predominantly on its size and heliocentric distance, independent of nucleus age, shape, local topography, and active fraction.

We suggest that cometary activity is largely restricted to steep slopes (above the angle of repose), where mass wasting events can maintain volatile ices exposed at the surface. Such mass wasting is likely to produce overlapping debris fields that produce fine layering of the comet surface.

Such a connection between mass wasting and comet activity suggests that the simple paradigm of comet activity slowly fading until extinction is oversimplified, and ignores the capability of mass wasting events to reactivate a dormant comet nucleus. We propose three theoretical regimes of mass wasting of comet nuclei, determined by a comparison of the interval between mass wasting events and the timescale of cometary activity fading: quasi-equilibrium, episodic, and quasi-dormant. Because these regimes are not permanent, we propose a new paradigm for the evolution of a comet nucleus's activity, in which erosion and activity fading tend to move comets over time into regimes of less frequent mass wasting events: from quasi-equilibrium to episodic activity, to quasi-dormant, and finally extinction of activity. However, the mass wasting events themselves can trigger sublimative torques and rotation state changes, which can either increase or decrease the frequency of future mass wasting events.

5. **Acknowledgements**

Jordan Steckloff would like to acknowledge Professor Jay Melosh, who provided support and funding for this research. We also with to thank M. Ramy El-Maarry and Jean-Baptiste Vincent, who greatly improved the quality of this paper through their reviews.